\documentclass[12pt]{article}
\usepackage{epsfig}
\usepackage{amsfonts}
\usepackage{latexsym}
\usepackage{amsmath}
\usepackage{mathrsfs}
\usepackage{hyperref}
\usepackage{setspace}
\usepackage{color}
\numberwithin{equation}{section}

\begin{document}

\begin{titlepage}
\unitlength = 1mm
%\today 
\begin{flushright}
KOBE-TH-14-12\\
QGASLAB-14-06
\end{flushright}

\vskip 1cm
\begin{center}

 {\Large {\textsc{\textbf{Entanglement negativity in the multiverse}}}}

\vspace{1.8cm}
Sugumi Kanno$^{*\,\flat\,\natural}$, Jonathan P. Shock$^{\natural\,\dag}$ and Jiro Soda$^\ddag$

\vspace{1cm}

{\it $^*$ Department of Theoretical Physics and History of Science,
University of the Basque Country UPV/EHU,
48080 Bilbao, Spain}

\vspace{0.2cm}

{\it $^\flat$ IKERBASQUE, Basque Foundation for Science, 
Maria Diaz de Haro 3,
48013, Bilbao, Spain}

\vspace{0.2cm}

{\it $^\natural$ Laboratory for Quantum Gravity \& Strings and Astrophysics, Cosmology \& Gravity Center,
Department of Mathematics \& Applied Mathematics, University of Cape Town,
Private Bag, Rondebosch 7701, South Africa}

\vspace{0.2cm}

{\it $^\dag$ National Institute for Theoretical Physics,
Private Bag X1,
Matieland, 7602, South Africa}

\vspace{0.2cm}

{\it $^\ddag$ Department of Physics, Kobe University, Kobe 657-8501, Japan}

\vskip 1.5cm

\begin{abstract}
\baselineskip=6mm
We explore quantum entanglement between two causally disconnected regions in the multiverse. We first consider a free massive scalar field, and compute the entanglement negativity between two causally separated open charts in de Sitter space. The qualitative feature of it turns out to be in agreement with that of the entanglement entropy. 
We then introduce two observers who determine the entanglement between two causally disconnected de Sitter spaces. When one of the observers remains constrained to a region of the open chart in a de Sitter space, we find that the scale dependence enters into the entanglement.
We show that a state which is initially maximally entangled becomes more entangled or less entangled on large scales depending on the mass of the scalar field and recovers the initial entanglement in the small scale limit. We argue that quantum entanglement may provide some evidence for the existence of the multiverse. 
\end{abstract}

\vspace{1.0cm}

\end{center}
\end{titlepage}

\pagestyle{plain}
\setcounter{page}{1}
\newcounter{bean}
\baselineskip18pt

\setcounter{tocdepth}{2}

\tableofcontents

\section{Introduction}

Historically, quantum entanglement has been one of the most fascinating but controversial features of quantum mechanics, since Einstein-Podolsky-Rosen (EPR) pointed out that performing a local measurement may affect the outcome of local measurements instantaneously beyond the lightcone in 1935~\cite{Einstein:1935rr}. %A remarkable theoretical effort has been devoted to classifying and formulating the measures of the quantum entanglement. 
It was Aspect et al's experiment that caused a paradigm shift in 1981. They performed a convincing test that the quantum entanglement is a fundamental aspect of quantum mechanics by measuring correlations of linear polarizations of pairs of photons ~\cite{Aspect:1981zz, Aspect:1982fx}. Since then, more interest has been paid to how to make use of quantum entanglement of EPR pairs in quantum cryptography and quantum teleportation~(see \cite{Horodecki:2009zz} and references therein). 

Entanglement entropy has proved to be a useful quantitative measure of entanglement of a quantum system. Nowadays, entanglement entropy has become a useful tool in understanding phenomena in condensed matter physics, quantum information and high energy physics. Nonetheless, the actual calculation of the entanglement entropy in quantum field theories has not been an easy task. Ryu and Takayanagi made great progress recently by discovering a powerful method of calculating the entanglement entropy of a strongly coupled quantum field theory with a gravity dual using holographic techniques~\cite{Ryu:2006bv}. Their formula has so far passed many consistency checks. 

In order to discuss the gravitational dual of the entanglement entropy in a quantum field theory in the Bunch-Davies vacuum in de Sitter space, Maldacena and Pimentel developed an explicit method in \cite{Maldacena:2012xp}. Their method was also extended to $\alpha$-vacua in \cite{Kanno:2014lma, Iizuka:2014rua}. The quantum entanglement could exist beyond the size of the Hubble horizon if a pair of particles created within causally connected Hubble horizon size region was separated off by the de Sitter expansion. In fact, this research showed that two causally disconnected regions in de Sitter space are entangled. This suggests that vacuum fluctuations observed in our universe may be entangled with those in another part of the multiverse. In fact, by using the reduced  density matrix derived in \cite{Maldacena:2012xp}, it was found that the quantum entanglement affects the shape of the spectrum on large scales comparable to or greater than the curvature radius in \cite{Kanno:2014ifa}. This could be an observational signature of the multiverse. 

In an era of precision cosmology, observational technologies may provide further understanding of our universe by making use of quantum entanglement. Indeed, there are some attempts to apply the effect of quantum entanglement to cosmology, with a variety of motivations and formalisms in \cite{Ball:2005xa, VerSteeg:2007xs, Fuentes:2010dt, Nambu:2011ae, Albrecht:2014aga, Lim:2014uea}. Also, the results of \cite{Garriga:2012qp, Garriga:2013pga, Frob:2014zka} suggest that the frame of bubble nucleation is observer dependent, determined by the rest frame of the observer, so the quantum entanglement produced in the process of bubble nucleation may be observer dependent as well. To reveal such an observer dependence of the quantum entanglement in the multiverse is one of our motivations of this work.

Another motivation is to understand the result of \cite{Kanno:2014ifa} better in the multiverse picture. Inflationary cosmology and the string landscape suggest that our universe may not be the only universe but part of a vast complex of universes that we call the multiverse \cite{Sato:1981gv, Vilenkin:1983xq, Linde:1986fc, Linde:1986fd, Bousso:2000xa, Susskind:2003kw}. Until recently, however, this multiverse idea has been criticized as a philosophical proposal that cannot be tested. However, there may be quantum entanglement between two causally separated universes in the multiverse, and it may produce detectable signatures as demonstrated in \cite{Kanno:2014ifa}.
In the structure of the multiverse, there may be many causally disconnected de Sitter universes (de Sitter bubbles). Some of their quantum states may be far from the Bunch-Davies vacuum and be entangled with those of the other part of the multiverse as shown in~\cite{Maldacena:2012xp}. To model such a situation, we consider two separated de Sitter spaces supposing that they are in a maximally entangled pure state initially.  
We then introduce two observers who determine the entanglement between the two causally disconnected de Sitter spaces. We assume that one of the observers is inside of a de Sitter universe (a de Sitter bubble) and want to see how the inside observer detects the signature of entanglement with another de Sitter universe (another de Sitter bubble).
Since the inside observer has no access to the region outside their universe, the observer must trace over the disconnected, inaccessible outside region and thus lose information about it. Then the observer's state is going to be a mixed state. On the other hand, the other observer  remains in a pure state of the other separated de Sitter space, so we need to consider the entanglement between the mixed and pure states. The entanglement entropy defined as the von Neumann entropy is a measure of entanglement for a biparticle pure state. For any mixed state of an arbitrary biparticle system, negativity or logarithmic negativity is known as a measure of entanglement~\cite{Vidal:2002zz, Plenio:2005, Calabrese:2012, Rangamani:2014ywa}. This measure would be useful for analyzing the entanglement between two observers of causally disconnected de Sitter universes.

In this paper, firstly, we calculate the negativity of quantum entanglement of a massive scalar field in a de Sitter background. More precisely, we compute the negativity for two causally disconnected open chart in de Sitter space and check the consistency of qualitative feature obtained by calculating the entanglement entropy in \cite{Maldacena:2012xp}. Next, we try to extend the quantum system by introducing two de Sitter spaces, with a hope to extract more information about quantum entanglement of the multiverse. We calculate the negativity of a quantum state with an initially maximally entangled state. We will see the entanglement becomes more or less entangled on large scales depending on the scalar mass and recovers the initial entangled state in the small scale limit.

The paper is organized as follows. In section \ref{s2} we review the necessary definitions of negativity and logarithmic negativity with some simple examples. In section \ref{s3}, we review the method developed in~\cite{Maldacena:2012xp} with some comments relevant to the calculation of the negativity. We then calculate the logarithmic negativity for two causally disconnected regions. In section \ref{s4}, we introduce two observers who determine the entanglement between the two causally disconnected de Sitter spaces, and calculate the logarithmic negativity between them. Our results are summarized and their implication is discussed in section \ref{s5}.

\section{Negativity and logarithmic negativity}
\label{s2}

To characterize the entanglement of a quantum state, there have been many entanglement measures proposed. The negativity is one such measure of quantum entanglement, which is derived from the positive partial transpose criterion for 
separability~\cite{Horodecki:2009zz}.
Here, we explain the definition of negativity and logarithmic negativity.

We consider a quantum mechanical system consisting of subspaces $A$ and $B$.  The Hilbert space becomes a direct product ${\cal H}={\cal H}_A\otimes{\cal H}_B$. 
For a pure state, we know any state has a Schmidt decomposition
\begin{eqnarray}
|\psi\rangle=\sum_i\sqrt{\lambda_i}\,|i\rangle_A \otimes |i\rangle_B\,,
\label{schmidt1}
\end{eqnarray}
where $\lambda_i$ is the probability to observe the $i$-th state and satisfies $\sum_i \lambda_i =1$.
In this case, the reduced density operator of the subsystem $A$ is calculated by tracing over the degrees of freedom of $B$ and is given by
\begin{eqnarray}
  \rho_A = {\rm Tr}_B \  |\psi\rangle \langle \psi |  = \sum_i \lambda_i \,|i\rangle_A \, {}_A\langle i|\,.
\end{eqnarray}
The entanglement entropy is defined via the density matrix as the von Neumann entropy
\begin{eqnarray}
  S = - {\rm Tr} \rho_A \log \rho_A = - \sum_i  \lambda_i \log \lambda_i\,.
\end{eqnarray}
When there is no entanglement, namely, $\lambda_1 =1$ and $ \lambda_{i\neq 1}=0$, the entanglement entropy vanishes. Therefore, the entanglement entropy is a good measure of the quantum entanglement.
However, the entanglement entropy also gives a nonzero value even in the presence of classical correlations that have the state mixed. In such a case, the entanglement entropy does not distinguish quantum correlations from classical ones~\cite{Horodecki:2009zz}. 

As a powerful measure in such cases, negativity and logarithmic negativity based on a criterion for separability is known. The idea is to characterize an entangled state as a state that is not separable. A state is separable if and only if the density operator of the total system is expressed as
a sum of tensor products of the density operator of subsystems:
\begin{eqnarray}\label{nonentangled}
\rho=\sum_i\lambda_i\,\rho^A_i\otimes\rho^B_i
=\sum_i\lambda_i\,|i\rangle_A\,{}_A\langle i|\otimes  |i\rangle_B\,{}_B\langle i|
\,,\qquad\qquad \lambda_i\geq 0\,,
\end{eqnarray}
where  $ \rho^A_i=|i\rangle_A\,{}_A\langle i| $,  $ \rho^B_i=|i\rangle_B\,{}_B\langle i| $, and $\rho$ is a density operator
of the total system.
If we consider a general density operator including entangled and non-entangled states, which is expanded as
\begin{eqnarray}
\rho=\sum_{i,j,k,\ell}C_{ijk\ell}\,|i\rangle_A\,{}_A\langle j|\otimes|k\rangle_B\,{}_B\langle\ell|\,,
\end{eqnarray}
with coefficients $C_{ijk\ell}$.
Taking a partial transpose with respect to the subsystem $A$, we obtain a new operator
\begin{eqnarray}
\rho^{T_A}=\sum_{i,j,k,\ell}C_{ijk\ell}\,|j\rangle_A\,{}_A\langle i|\otimes|k\rangle_B\,{}_B\langle\ell|\,.
\end{eqnarray}
On the other hand, for non-entangled state Eq.~(\ref{nonentangled}), its partial transpose is unchanged
\begin{eqnarray}
\rho^{T_A}=\sum_i\lambda_i\,|i\rangle_A\,{}_A\langle i|\otimes  |i\rangle_B\,{}_B\langle i|\,.
\end{eqnarray}
Thus, we find $\rho^{T_A}\geq 0$ (ie. all eigenvalues are positive) for the non-entangled state. In other words, if $\rho^{T_A}$ has a negative eigenvalue, 
the density operator $\rho$ cannot be written as Eq.~(\ref{nonentangled}) and the state is guaranteed to be entangled.

Thus, we come to the following definition of an entanglement measure called negativity.
The negativity is defined by summing over all the negative eigenvalues
\begin{eqnarray}
{\cal N}=\sum_{\lambda_i<0}|\lambda_i|\,.
\end{eqnarray}
Thus, it would seem that when ${\cal N}=0$, there exists no entanglement.
However, unfortunately, this measure is not additive and not suitable for multi-subsystems. Hence, we need to define another entanglement measure named the logarithmic negativity as an improvement on the entanglement negativity.
To this end, we introduce the trace norm of $\rho^{T_A}$:
\begin{eqnarray}
\parallel\rho^{T_A}\parallel=\sum_{i}|\lambda_i|=\sum_{\lambda_i>0}\lambda_i
+\sum_{\lambda_i<0}|\lambda_i|\,,
\end{eqnarray}
where the trace norm is defined by $\parallel X\parallel={\rm Tr}\sqrt{X^\dagger X}$ of the operator $X$. We find it written in terms of the negativity as
\begin{eqnarray}
\parallel\rho^{T_A}\parallel=2\sum_{\lambda_i<0}|\lambda_i|+1=2{\cal N}+1\,,
\end{eqnarray}
due to the conservation of probability ${\rm Tr}\rho=1$ and ${\rm Tr}\rho^{T_A}=1$, and thus $\Sigma_i\,\lambda_i=1$. Then the negativity can written as
\begin{eqnarray}
{\cal N}=\frac{\parallel\rho^{T_A}\parallel-1}{2}\,.
\end{eqnarray}
The logarithmic negativity is then defined as
\begin{eqnarray}
L{\cal N}=\log\parallel\rho^{T_A}\parallel=\log\left(2{\cal N}+1\right)\,.
\label{lnvsn}
\end{eqnarray}
When $L{\cal N}\neq 0$, the state is entangled.

Let us illustrate how to compute the negativity for a pure state.  We can use the Schmidt decomposition for a pure state 
\begin{eqnarray}
|\psi\rangle=\sum_i\sqrt{\lambda_i}\,|i\rangle_A\otimes |i\rangle_B\,.
\label{schmidt}
\end{eqnarray}
Then the partial transpose of the density matrix $\rho=|\psi\rangle\langle\psi|$ is given by
\begin{eqnarray}
\rho^{T_A}&=&\sum_{i,j}\sqrt{\lambda_i\lambda_j}\,\left(\,|i\rangle_A\otimes|i\rangle_B\,{}_A\langle j|\otimes{}_B\langle j|\,\right)^{T_A}\nonumber\\
&=&\sum_{i,j}\sqrt{\lambda_i\lambda_j}\,|j\rangle_A\,{}_A\langle i|\otimes|i\rangle_B\,{}_B\langle j|\nonumber\\
&=&\sum_{i}\lambda_i\,|i\rangle_A\,{}_A\langle i|\otimes|i\rangle_B\,{}_B\langle i|+
\sum_{i\neq j}\sqrt{\lambda_i\lambda_j}\,|j\rangle_A\,{}_A\langle i|\otimes|i\rangle_B\,{}_B\langle j|\,.
\label{pta}
\end{eqnarray}
If we introduce
\begin{eqnarray}
|\psi^\pm_{ij}\rangle=\frac{1}{\sqrt{2}}\left(\,|i\rangle_A\otimes|j\rangle_B\pm
|j\rangle_A\otimes|i\rangle_B\,\right)\qquad\qquad {\rm for}\quad i<j\,,
\end{eqnarray}
we find that $|\psi^\pm_{ij}\rangle$ is orthogonal to $|i\rangle_A\otimes|i\rangle_B$. Then the partial transpose of $\rho$ in Eq.~(\ref{pta}) is written as $|\psi^\pm_{ij}\rangle$ and found to be diagonalized as
\begin{eqnarray}
\rho^{T_A}&=&\sum_{i}\lambda_i\,|i\rangle_A\,{}_A\langle i|\otimes|i\rangle_B\,{}_B\langle i|+
\sum_{i<j}\sqrt{\lambda_i\lambda_j}\,\left(\,|\psi^+_{ij}\rangle\langle\psi^+_{ij}|
-|\psi^-_{ij}\rangle\langle\psi^-_{ij}|\,\right)\,.
\label{diagonal}
\end{eqnarray}
We can read off the eigenvalues of the partial transpose of $\rho$ as follows
\begin{eqnarray}
\lambda_i,\quad\sqrt{\lambda_i\lambda_j},\quad-\sqrt{\lambda_i\lambda_j}\,.
\end{eqnarray}
We see that negative eigenvalues $-\sqrt{\lambda_i\lambda_j}$ exist. When there exists a negative eigenvalue, the state $|\psi\rangle$ is entangled. Note that if at least two of the $\lambda_i$ ($i>2$) aren't zero, the negative eigenvalues always exist.
From the diagonalized form  Eq.~(\ref{diagonal}), we find
\begin{eqnarray}
\parallel\rho^{T_A}\parallel=\sum_{i}\lambda_i+2\sum_{i<j}\sqrt{\lambda_i\lambda_j}
=\sum_{i,j}\sqrt{\lambda_i\lambda_j}=\left(\sum_i\sqrt{\lambda_i}\right)^2\,.
\end{eqnarray}
Then the logarithmic negativity is calculated to be
\begin{eqnarray}
L{\cal N}=2\log\left(\sum_i\sqrt{\lambda_i}\right)\,.
\label{logneg}
\end{eqnarray}
For a $d$-dimensional maximally entangled state $|\psi\rangle$, $\lambda_i=1/d$. Then the logarithmic negativity is found to be
\begin{eqnarray}
L{\cal N}=2\log\left(\sum_i\frac{1}{\sqrt{d}}\right)=\log d\,.
\end{eqnarray}

For general quantum states such as mixed states, we need to calculate the negativity or logarithmic negativity numerically. In the next section, we shall apply this measure to quantify the entanglement of a quantum state in de Sitter space.

\section{Negativity between two causally disconnected open charts}
\label{s3}

\begin{figure}[t]
\vspace{-3cm}
\includegraphics[height=11cm]{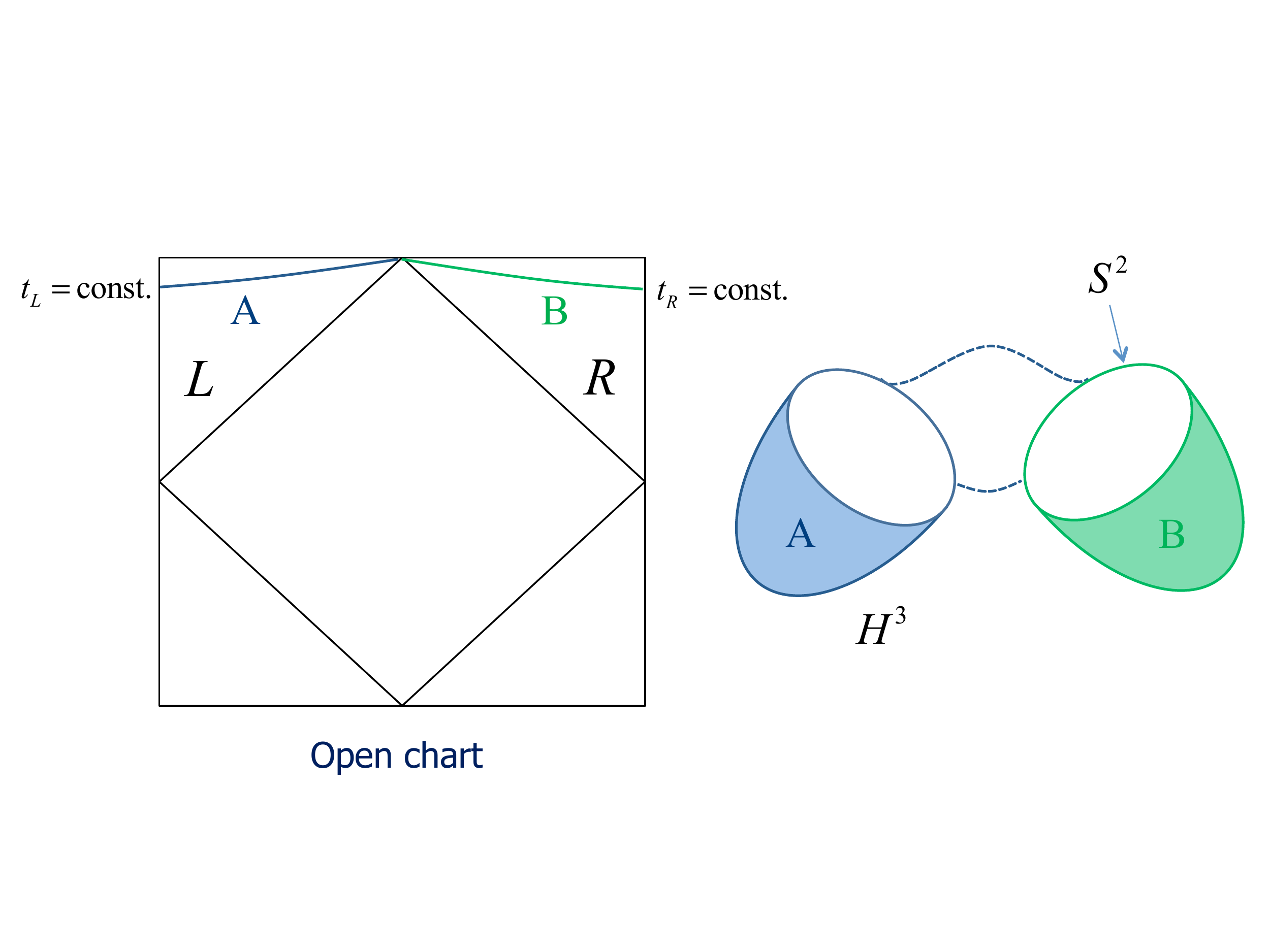}\centering
\vspace{-1cm}
\caption{The Penrose diagram of the open chart is shown. The spatial hypersurface is depicted on the right.}
\label{fig1}
\end{figure}

In this section, we first review the derivation of the reduced density matrix in the open chart developed by \cite{Maldacena:2012xp} and then compute the logarithmic negativity between two causally disconnected open charts in de Sitter space. The Penrose diagram of the open chart is given in Figure~\ref{fig1}. The open chart of de Sitter space is studied in detail in \cite{Sasaki:1994yt}. It is known that the inside of a nucleated bubble looks like an open universe \cite{Coleman:1980aw}, so this formulation will be suitable for the multiverse framework.

\subsection{Mode functions in the open chart}
We consider a free massive scalar field in de Sitter space with the action given by
\begin{eqnarray}
S=\int d^4 x \sqrt{-g} \left[\, -\frac{1}{2}\,g^{\mu\nu}\partial_\mu\phi\,\partial_\nu \phi
-\frac{m^2}{2}\phi^2\,\right]\,.
\label{action}
\end{eqnarray}
The metric in each $R$ and $L$ region is obtained by analytic continuation 
from a Euclidean four-sphere metric and expressed, respectively, as
\begin{eqnarray}
ds^2_R&=&H^{-2}\left[-dt^2_R+\sinh^2t_R\left(dr^2_R+\sinh^2r_R\,d\Omega^2\right)
\right]\,,\nonumber\\
ds^2_L&=&H^{-2}\left[-dt^2_L+\sinh^2t_L\left(dr^2_L+\sinh^2r_L\,d\Omega^2\right)
\right]\,,
\end{eqnarray}
where $d\Omega^2$ is the metric on the two-sphere. Because the $R$ and $L$ regions are completely symmetric, we write $(t,r)=(t_R, r_R)$ or $(t_L, r_L)$. If we perform the separation of variables,
\begin{eqnarray}
\phi=  \frac{H}{\sinh t}\,  \chi_p (t) \,Y_{p\ell m}(r,\Omega)\,,
\end{eqnarray} 
the equations of motion for $\chi_p$ and $Y_{p\ell m}$ in the $R$ or $L$ regions are found to be in common
\begin{eqnarray}
&&\left[\,\frac{\partial^2}{\partial t^2}+3\coth t\,\frac{\partial}{\partial t}
+\frac{1+p^2}{\sinh^2t}+\frac{m^2}{H^2}\right]\chi_p(t)=0\,,
\label{eom}\\
&&\left[\,\frac{\partial^2}{\partial r^2}+2\coth r\,\frac{\partial}{\partial r}
-\frac{1}{\sinh^2r}\,\rm\bf L^2\,\right]Y_{p\ell m}(r,\Omega)=-(1+p^2)Y_{p\ell m}(r,\Omega)\,,
\end{eqnarray}
where $\rm\bf L^2$ is the Laplacian operator on the unit two-sphere, and $Y_{p\ell m}$ are harmonic functions on the 
three-dimensional hyperbolic space. 
It is natural to choose the initial state to be the de Sitter invariant vacuum, so we consider the positive frequency mode functions corresponding to the Bunch-Davies vacuum. Then the time dependent part of $\chi_p(t)$ is found to be
\begin{eqnarray}
\chi_{p,R}(t)&=&\frac{1}{N_b}\left( P_{\nu-1/2}^{ip}(\cosh t_R)+
\frac{\cos\pi\nu}{i\sinh\pi p}P_{\nu-1/2}^{ip}(\cosh t_L) \right.\nonumber\\
 && \left. \qquad\qquad
 -\frac{\cos\left(ip+\nu\right)\pi}{i\sinh\pi p}\,e^{-\pi p}\,\frac{\Gamma(\nu+ip+1/2)}{\Gamma(\nu-ip+1/2)}P_{\nu-1/2}^{-ip}(\cosh t_L)
\right)\,,
\label{chi1}
\end{eqnarray}
\begin{eqnarray}
\chi_{p,L}(t)&=&\frac{1}{N_b}\left(P_{\nu-1/2}^{ip}(\cosh t_L)+
\frac{\cos\pi\nu}{i\sinh\pi p}P_{\nu-1/2}^{ip}(\cosh t_R) \right.\nonumber\\
&& \left.\qquad\qquad
 -\frac{\cos\left(ip+\nu\right)\pi}{i\sinh\pi p}\,e^{-\pi p}\,\frac{\Gamma(\nu+ip+1/2)}{\Gamma(\nu-ip+1/2)}P_{\nu-1/2}^{-ip}(\cosh t_R)
\right)\,,
\label{chi11}
\end{eqnarray}
where $P^{\pm ip}_{\nu-\frac{1}{2}}$ are the associated Legendre functions and $N_b$ is a normalization factor expressed as
\begin{eqnarray}
N_b^2=\frac{2}{\pi}e^{-\pi p}
(\cosh2\pi p+\cos2\pi\nu)\,.
\end{eqnarray}
We have defined a mass parameter
\begin{eqnarray}
\nu=\sqrt{\frac{9}{4}-\frac{m^2}{H^2}}\,.
\end{eqnarray}
When $\nu=1/2$ we have a conformally coupled scalar and the system is conformally invariant. 
The massless case corresponds to $\nu=3/2$. Note that (\ref{chi1}) and (\ref{chi11}) are two independent solutions for the entire de Sitter space. The solution (\ref{chi1}) (or (\ref{chi11})) is expressed by the linear combination of the positive frequency function in the $R$ (or $L$) region and the analytic continuation of it to the $L$ (or $R$) region.\footnote{The expression of the solutions is different from the symmetrized form in~\cite{Sasaki:1994yt} but the resultant reduced density matrix is the same as we will see in Eq.~(\ref{dm}). See \cite{Maldacena:2012xp, Kanno:2014lma, Kanno:2014ifa} for the reduced density matrix obtained by the symmetrized form.}

We expand the field in terms of the creation and annihilation operators
\begin{eqnarray}
\hat\phi(t,r,\Omega) &=& \frac{H}{\sinh t}\int dp \sum_{\sigma,\ell,m} 
\left[\,a_{\sigma p\ell m}\,\chi_{p,\sigma}(t)
+a_{\sigma p\ell -m}^\dagger\,\chi^*_{p,\sigma}(t)\,\right]Y_{p\ell m}(r,\Omega)
\nonumber\\
&=&\frac{H}{\sinh t}\int dp \sum_{\ell,m}\phi_{p\ell m}(t)Y_{p\ell m}(r,\Omega)
\,,
\end{eqnarray}
where $\sigma=(R,L)$, and without loss of generality, we assumed the normalization of $Y_{p\ell m}$ is such that $Y_{p\ell m}^*=Y_{p\ell-m}$. We introduced a Fourier mode field operator
\begin{eqnarray}
\phi_{p\ell m}(t)\equiv
\sum_\sigma\left[\,a_{\sigma p\ell m}\,\chi_{p,\sigma}(t)
+a_{\sigma p\ell -m}^\dagger\,\chi^*_{p,\sigma}(t)\right]\,,
\label{phi1}
\end{eqnarray}
where $a_{\sigma p\ell m}$ satisfies $a_{\sigma p\ell m}|{\rm BD}\rangle=0$.

For convenience, we write the mode functions and the associated Legendre functions of the $R$ and $L$ regions in a simple form
 $\chi_{p,R,L}(t)\equiv\chi^{R, L}\,,P^{R, L}\equiv P_{\nu-1/2}^{ip}(\cosh t_{R,L})\,,\,P^{R*, L*}\equiv P_{\nu-1/2}^{-ip}(\cosh t_{R,L})$.
 Also we omit the indices $p, \ell, m$ of $\phi_{p\ell m}$ and $a_{p\ell m}$ for simplicity unless there may be any confusion below.

\subsection{The reduced density matrix}
Now we introduce the mode functions in each $R$ or $L$ region, which are given by
\begin{eqnarray}
\varphi^q=\tilde{N}_b^{-1}P^q\,,\qquad \tilde{N}_b=\frac{\sqrt{2p}}{|\Gamma(1+ip)|}\,,
\end{eqnarray}
where $q=(R, L)$. Let us introduce vectors with four components
\begin{eqnarray}
\chi^I=\left(
\begin{array}{l}
\chi^\sigma\\
\chi^{\sigma*}
\end{array}\right)\,,\qquad
\varphi^J=\left(
\begin{array}{l}
\varphi^q\\
\varphi^{q*}
\end{array}\right)\,,
\label{M}
\end{eqnarray}
and a $4\times 4$ matrix
\begin{eqnarray}
M^I{}_J=\left(
\begin{array}{ll}
\alpha^\sigma{}_{\!q} &~ \beta^\sigma{}_{\!q} \vspace{3mm}\\
\beta^{\sigma*}{}_{\!\!\!q} &~ \alpha^{\sigma*}{}_{\!\!\!q} \\
\end{array}\right)\,,
\label{M}
\end{eqnarray}
where $\sigma=(R, L)$. Here, we defined
\begin{eqnarray}
\alpha^\sigma{}_{\!q}=\left(
\begin{array}{ll}
A &~ B \vspace{3mm}\\
B &~ A \\
\end{array}\right)\,,\qquad
\beta^\sigma{}_{\!q}=\left(
\begin{array}{ll}
\,0 & D \vspace{3mm}\\
D &\, 0 \\
\end{array}\right)\,,
\end{eqnarray}
with
\begin{eqnarray}
A=\frac{\tilde{N}_b}{N_b}\,,\qquad 
B=\frac{\tilde{N}_b}{N_b}\frac{\cos\pi\nu}{i\sinh\pi p}\,,\qquad 
D=-\frac{\tilde{N}_b}{N_b}\frac{\cos\left(ip+\nu\right)\pi}{i\sinh\pi p}\,e^{-\pi p}\,\frac{\Gamma(\nu+ip+1/2)}{\Gamma(\nu-ip+1/2)} \ .
\label{explicit}
\end{eqnarray}
Then the mode functions ~(\ref{chi1}) and (\ref{chi11}) are written as
\begin{eqnarray}
\chi^I=M^I{}_J\,\varphi^J\,.
\label{chi2}
\end{eqnarray}
Note that this procedure of changing the mode functions from $\chi^I$ to $\varphi^I$ is 
a Bogoliubov transformation. The Bogoliubov coefficients are given 
in terms of $\alpha$ and $\beta$ in the matrix $M$ in Eq.~(\ref{M}). 
Now let us introduce new creation and anihilation operators $b_I$ defined 
such that $b_R|R\rangle=0$ and $b_L|L\rangle=0$.

The Fourier mode field operator in Eq.~(\ref{phi1}) is now expressed as
\begin{eqnarray}
\phi(t)=a_I\,\chi^I=b_I\,\varphi^I\,,\qquad
a_{I}=\left(\,a_q\,,\,a_q^\dagger\,\right)\,,\qquad
b_{J}=\left(\,b_q\,,\,b_q^\dagger\,\right)\,.
\end{eqnarray}
Plugging Eq.~(\ref{chi2}) into the above, we find the relation between the operators $a_I$ and $b_I$ has to be
\begin{eqnarray}
a_J=b_I\left(M^{-1}\right)^I{}_J\,,\qquad
\left(M^{-1}\right)^I{}_J=\left(
\begin{array}{ll}
\xi^q{}_{\sigma} &~ \delta^q{}_{\sigma} \vspace{3mm}\\
\delta^{q*}{}_{\!\!\sigma} &~ \xi^{q*}{}_{\!\!\sigma} \\
\end{array}\right)\,,
\label{ab0}
\end{eqnarray}
where the components of the matrix $M^{-1}$ are calculated as
\begin{eqnarray}
\xi=
\left(\alpha-\beta\,\alpha^{*\,-1}\beta^*\right)^{-1}=\alpha^*\,,\qquad\quad
\delta=-\alpha^{-1}\beta\,\xi^*=-\beta\,.
\label{ab}
\end{eqnarray}
where we used the relations (\ref{explicit}).
Thus, we can regard the Bunch-Davies vacuum as a Bogoliubov transformation 
of the $R,L$-vacua as
\begin{eqnarray}
|{\rm BD}\rangle\propto\exp\left(\frac{1}{2}\sum_{i,j=R,L}m_{ij}\,b_i^\dagger\, b_j^\dagger\right) |R\rangle|L\rangle\,,
\label{bogoliubov1}
\end{eqnarray}
where $m_{ij}$ is a symmetric matrix and the operators $b_i$ satisfy the commutation relation $[b_i,b_j^\dagger]=\delta_{ij}$. We note that the normalization of the Bogoliubov transformation is omitted here because another Bogoliubov transformation will be used to derive the reduced density matrix. The condition $a_q|{\rm BD}\rangle=0$ determines $m_{ij}$:
\begin{eqnarray}
m_{ij}=-\delta_{i\sigma}^*\left(\xi^{-1}\right)_{\sigma j}
=e^{i\theta}\frac{\sqrt{2}\,e^{-p\pi}}{\sqrt{\cosh 2\pi p+\cos 2\pi\nu}}
\left(
\begin{array}{cc}
\cos \pi\nu & i\sinh p\pi \vspace{1mm}\\
i\sinh p\pi & \cos \pi\nu \\
\end{array}
\right)\,,
\label{mij}
\end{eqnarray}
where $e^{i\theta}$ contains all unimportant phase factors for $\nu^2>0$. 

In the cases of conformal invariance ($\nu=1/2$) and masslessness ($\nu=3/2)$, we see that the density matrix $\rho=|{\rm BD}\rangle\langle{\rm BD}|$ is going to be diagonal in the $|R\rangle|L\rangle$ basis when the state is written in the form of Eq.~(\ref{bogoliubov1}). In other cases, however, it is not diagonal and then it is difficult to trace over the $R$ degrees of freedom.
Thus, we perform a further Bogoliubov transformation by introducing new operators $c_R$ and $c_L$
\begin{eqnarray}
c_R = u\,b_R + v\,b_R^\dagger \,,\qquad\quad
c_L = \bar{u}\,b_L + \bar{v}\,b_L^\dagger\,,
\label{bc}
\end{eqnarray}
to obtain the relation
\begin{eqnarray}
|{\rm BD}\rangle = N_{\gamma_p}^{-1}
\exp\left(\gamma_p\,c_R^\dagger\,c_L^\dagger\,\right)|R'\rangle|L'\rangle\,.
\label{bogoliubov2}
\end{eqnarray}
Note that the normalizations $|u|^2-|v|^2=1$ and $|\bar{u}|^2-|\bar{v}|^2=1$ are assumed
so that the new operators satisfy the commutation relation 
$[c_i,c_j^\dagger]=\delta_{ij}$. 
The normalization factor $N_{\gamma_p}$ is given by
\begin{eqnarray}
N_{\gamma_p}^2
=\left|\exp\left(\gamma_p\,c_R^\dagger\,c_L^\dagger\,\right)|R'\rangle|L'\rangle
\right|^2
=\frac{1}{1-|\gamma_p|^2}\,,
\label{norm2}
\end{eqnarray}
where $|\gamma_p|<1$ should be satisfied.
Notice that the basis vacuum changes from $|R\rangle|L\rangle$ to 
$|R^\prime\rangle|L^\prime\rangle$ but this Bogoliubov transformation does not mix $R$ and $L$ Hilbert spaces because Eq.~(\ref{bc}) is a linear transformation between $c_q$ and $b_q$.
The consistency conditions for Eq.~(\ref{bogoliubov2}) are
\begin{eqnarray}
c_R\,|{\rm BD}\rangle= \gamma_p\,c_L^\dagger\,|{\rm BD}\rangle \,,\qquad 
c_L\,|{\rm BD}\rangle = \gamma_p\,c_R^\dagger\,|{\rm BD}\rangle\,.
\label{consistency}
\end{eqnarray}
If we write $m_{RR} = m_{LL}\equiv \omega$ and $m_{LR}=m_{RL}\equiv \zeta$
in Eq.~(\ref{mij}), we find that $\omega$ is real and $\zeta$ is pure imaginary 
for positive $\nu^2$. By inserting Eqs.~(\ref{bogoliubov1}) and (\ref{bc}) 
into Eq.~(\ref{consistency}), we get a system of four homogeneous equations
\begin{eqnarray}
&&\omega\,u + v -\gamma_p\,\zeta\,\bar{v}^* =0 \ , \qquad 
\zeta\,u - \gamma_p\,\bar{u}^* - \gamma_p\,\omega\,\bar{v}^* =0\,,
\label{system1}\\
&&\omega\,\bar{u} + \bar{v} -\gamma_p\,\zeta\,v^* =0 \ , \qquad
\zeta\,\bar{u} - \gamma_p\,u^* - \gamma_p\,\omega\,v^* =0\,,
\label{system2}
\end{eqnarray}
where $\omega^*=\omega$ and $\zeta^*=-\zeta$.
We see that setting $v^* =\bar{v}$ and $u^* =\bar{u}$ is possible 
if $\gamma_p$ is purely imaginary $\gamma_p^*=-\gamma_p$. 
This is always possible by adjusting the phase of $c_q$. Then we find that Eq.~(\ref{system2}) becomes identical with Eq.~(\ref{system1}) and 
the system is reduced to that of two homogeneous equations. 
We look for such $\gamma_p$, keeping the normalization condition 
$|u|^2-|v|^2=1$ satisfied. 

As a non-trivial solution in the system of equations (\ref{system1}),
 $\gamma_p$ must be
\begin{eqnarray}
\gamma_p=\frac{1}{2\zeta}\left[-\omega^2+\zeta^2+1-\sqrt{\left(\omega^2-\zeta^2-1\right)^2-4\zeta^2}\,\right]\,,
\label{gammap}
\end{eqnarray}
where we took a minus sign in front of the square root term to 
satisfy $|\gamma_p|<1$. Note that $\gamma_p$ is purely imaginary.
Putting the $\omega$ and $\zeta$ defined in Eq.~(\ref{mij}) 
into Eq.~(\ref{gammap}), we obtain 
\begin{eqnarray}
\gamma_p = i\frac{\sqrt{2}}{\sqrt{\cosh 2\pi p + \cos 2\pi \nu}
 + \sqrt{\cosh 2\pi p + \cos 2\pi \nu +2 }}\,.
\label{gammap2}
\end{eqnarray}

Also $\bar{u}$, $\bar{v}$ are obtained by solving Eq.~(\ref{system1}) with the 
solution Eq.~(\ref{gammap2}) and imposing the normalization condition 
$|\bar{u}|^2-|\bar{v}|^2=1$, that is
\begin{eqnarray}
\bar{u}=\frac{1-\gamma_p\zeta}{\sqrt{|1-\gamma_p\zeta|^2-|\omega|^2}}
\,,\qquad
\bar{v}=\frac{\omega}{\sqrt{|1-\gamma_p\zeta|^2-|\omega|^2}}\,.
\end{eqnarray}
We find that $u$ and $v$ are real, that is, $u=u^*=\bar{u}$ and $v=v^*=\bar{v}$. 
Note that the phase factors for $u$ and $v$ are unimportant because they are canceled out in Eq.~(\ref{bogoliubov2}) by adjusting the phase of $c_q$.

Finally, we have the density matrix in the diagonalized form. By using  Eqs.~(\ref{bogoliubov2}) and (\ref{norm2}), the reduced density 
matrix is then found to be
\begin{eqnarray}
\rho_L ={\rm Tr}_{R}\,|{\rm BD}\rangle\langle{\rm BD}| 
=\left(1-|\gamma_p|^2\,\right)\sum_{n=0}^\infty 
|\gamma_p |^{2n}\,|n;p\ell m\rangle\langle n;p\ell m|\,,
\label{dm} 
\end{eqnarray}
where we defined $|n;p\ell m\rangle=1/\sqrt{n!}\,(c_L^\dagger)^n\,|L'\rangle$. 
Note that this density matrix is for each mode labeled by $p,\ell, m$.

In the cases of conformal invariance ($\nu=1/2$) and masslessness ($\nu=3/2$), we find $\gamma_p=e^{-\pi p}$. Then the reduced density matrix is given by
\begin{eqnarray}
\rho_L=\left(1-e^{-2\pi p}\right)\sum_{n=0}^\infty e^{-2\pi pn}\,|n;p\ell m\rangle\langle n;p\ell m|\,.
\label{thermal}
\end{eqnarray}
The resulting Eq.~(\ref{thermal}) is a thermal state with temperature
\begin{eqnarray}
T=\frac{H}{2\pi}\,.
\end{eqnarray}

\subsection{Negativity between the regions $R$ and $L$}
\label{s3.3}

In this subsection, we compute the entanglement negativity between two causally disconnected regions $R$ and $L$ in de Sitter space and compare it with the result of entanglement entropy calculated in \cite{Maldacena:2012xp}. 

\begin{figure}[t]
\begin{center}
\vspace{-1cm}
\begin{minipage}{8.1cm}
\includegraphics[height=6cm]{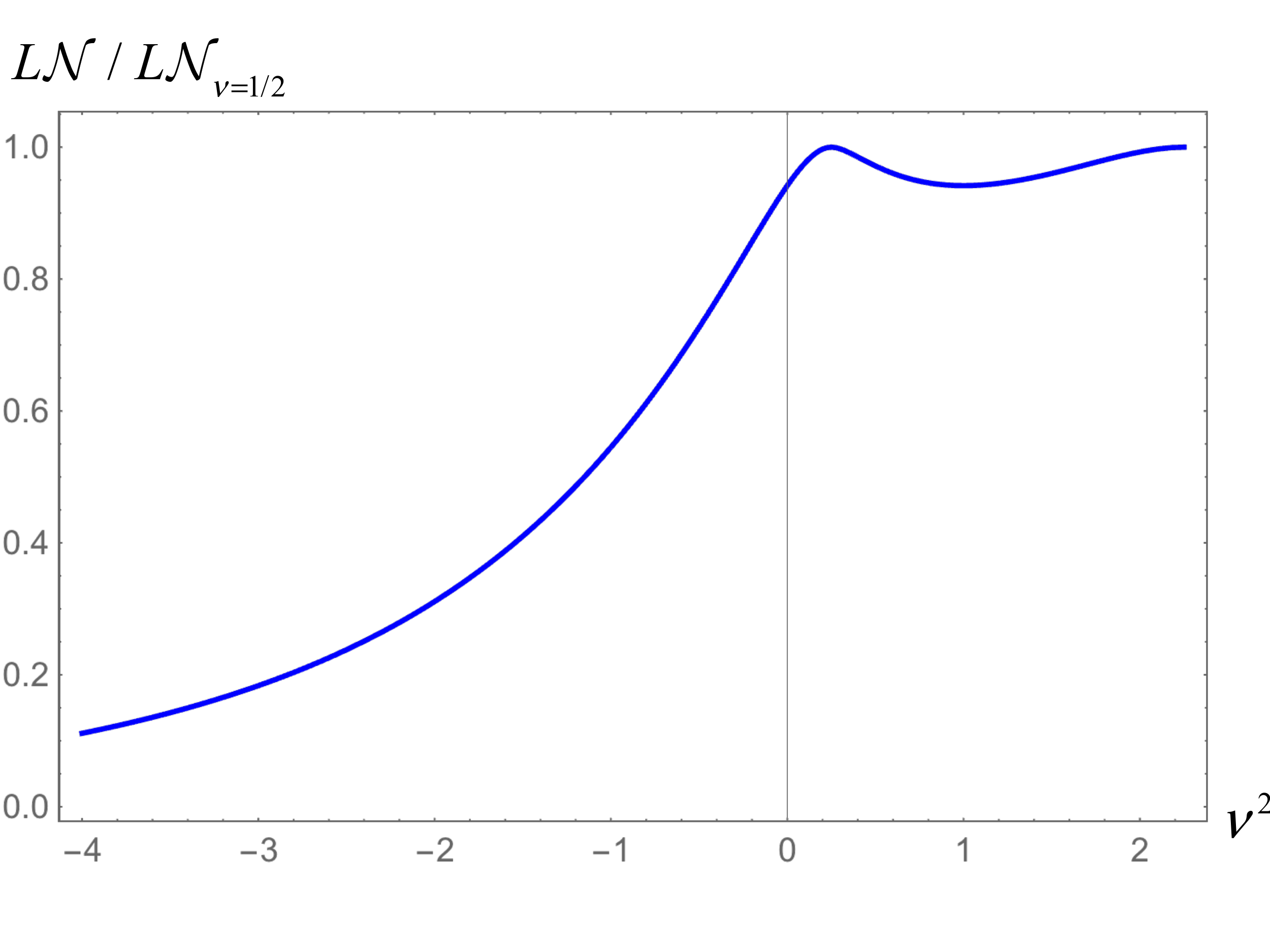}\centering
\end{minipage}
\begin{minipage}{8.1cm}
\hspace{0.5cm}
\includegraphics[height=6cm]{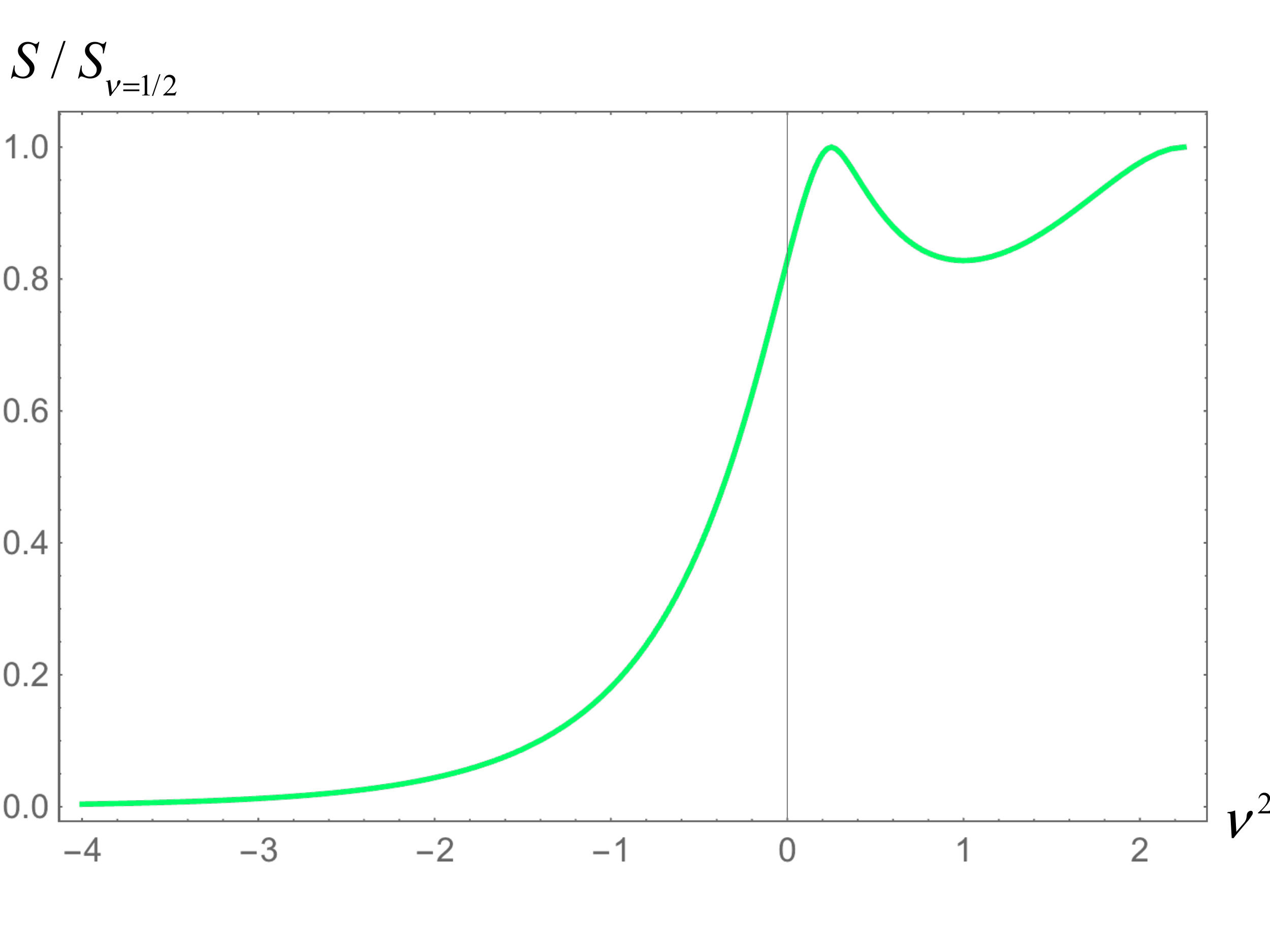}
\end{minipage}
\caption{Plots of the logarithmic negativity $L{\cal N}/{L{\cal N}_{\nu=1/2}}$ (Left) and the entanglement entropy $S/S_{\nu=1/2}$ (Right)  of the free massive scalar field, normalized to the conformally coupled scalar, versus its mass parameter squared. The massless case corresponds to $\nu^2=9/4$, the conformally coupled scalar to $\nu^2=1/4$. The qualitative features agree with each other and for large mass (negative $\nu^2$) both the entanglement entropy and the negativity decay exponentially.}
\label{fig2}
\end{center}
\end{figure}

From the previous subsection, we find the Bunch-Davies state in de Sitter space is given by Eqs.~(\ref{bogoliubov2}) and (\ref{norm2}). 
\begin{eqnarray}
|\rm BD\rangle&=&\sqrt{1-|\gamma_p|^2}\,\exp\left(\gamma_p\,c_R^\dagger\, c_L^\dagger\right)|R^\prime\rangle|L^\prime\rangle\nonumber\\
&=&\sqrt{1-|\gamma_p|^2}\,\sum_{n=0}^\infty\gamma_p^n\,|n;p\ell m\rangle_{R^\prime}|n;p\ell m\rangle_{L^\prime}\,,
\label{schmidt2}
\end{eqnarray}
where the states $|n;p\ell m\rangle_{R^\prime}$ and $|n;p\ell m\rangle_{L^\prime}$ are $n$ particle excitation states in $R^\prime$ and $L^\prime$-vacua.
Remembering the Schmidt decomposition for a pure state Eq.~(\ref{schmidt}), we can read off the corresponding eigenvalues
\begin{eqnarray}
\sqrt{\lambda_i}=\sqrt{1-|\gamma_p|^2}\,|\gamma_p|^n\,.
\end{eqnarray}
Then the logarithmic negativity in Eq.~(\ref{logneg}) for each mode is found to be
\begin{eqnarray}
L{\cal N}(p,\nu)=2\log\left(\sum_n\sqrt{1-|\gamma_p|^2}\,|\gamma_p|^n\right)
=\log\frac{1+|\gamma_p|}{1-|\gamma_p|}\,.
\end{eqnarray}
From Eq.~(\ref{lnvsn}), the negativity is
\begin{eqnarray}
{\cal N}(p,\nu)=\frac{|\gamma_p|}{1-|\gamma_p|}\,.
\end{eqnarray}
Since $|\gamma_p|\neq 0$ for a finite $p$, we find that the regions $R$ and $L$ are entangled~\footnote{We also find that the entanglement enhances on large scales when the scalar field is the cases of masslessness or conformal invariance because $|\gamma_p|\rightarrow 1$ as $p\rightarrow 0$ for $\nu=1/2,\,3/2$, which is consistent with the result of \cite{Kanno:2014ifa}.}.
Then the entanglement negativity between two causally disconnected regions $R$ and $L$ are obtained by integrating over $p$ and a volume integral over the hyperboloid,
\begin{eqnarray}
L{\cal N}(\nu)=\frac{1}{\pi}\int_0^\infty dp\,p^2L{\cal N}(p,\nu)\,.
\end{eqnarray}
The result normalized to the conformally coupled scalar ($\nu=1/2$) is plotted in the left panel of Figure~\ref{fig2}. 

Now let us compare the result with the entanglement entropy. 
The entanglement entropy for each mode derived by~\cite{Maldacena:2012xp} was expressed as
\begin{eqnarray}
S(p,\nu)=-{\rm Tr}\,\rho_L(p)\,{\rm log}\,\rho_L(p)
=-{\rm log}\,\left(1-|\gamma_p|^2\right)
-\frac{|\gamma_p|^2}{1-|\gamma_p|^2}\,{\rm log}\,|\gamma_p|^2\,.
\label{s}
\end{eqnarray}
Then the entanglement entropy between two causally disconnected regions are obtained by integrating over $p$ and a volume integral over the hyperboloid,
\begin{eqnarray}
S(\nu)=\frac{1}{\pi}\int_0^\infty dp\,p^2S(p,\nu)\,.
\end{eqnarray}
This is plotted in the right panel of Figure~\ref{fig2}. We see that the qualitative features of them are consistent. For quantitative aspects as a measure of quantum entanglement, the entanglement entropy appears to quantify the entanglement more stringently.

\section{Negativity between two causally disconnected de Sitter spaces}
\label{s4}

In this section, we introduce two observers who determine the entanglement between two causally separated de Sitter spaces, supposing that they are in a maximally entangled pure state initially in the multiverse. 

Since $R$ and $L$ regions in de Sitter space are completely symmetric, we set one of the observers in, say, the $L$ region of a de Sitter space which is inside of a de Sitter universe (de Sitter bubble). We simplify things here by ignoring the bubble wall.  We assume that the other observer is in the global chart of the other de Sitter space. Now, we want to see how the inside observer detects the signature of entanglement with another de Sitter universe (another de Sitter bubble). Since the inside observer has no access to the causally disconnected $R$ region, the observer must trace over the $R$ region and thus lose information about the state of the inaccessible region. Then the observer's state is going to be a mixed state. On the other hand, the other observer remains in a pure state of the other separated de Sitter space, so we need to consider the entanglement between the mixed and pure states. The entanglement entropy defined as the von Neumann entropy is a measure of entanglement between two subsystems for a pure state. If any mixed state is involved in subsystems, we need to use negativity or logarithmic negativity as a measure of entanglement~\cite{Vidal:2002zz, Plenio:2005, Calabrese:2012, Rangamani:2014ywa}. This measure would be useful for analyzing the entanglement between two observers of causally disconnected de Sitter universes.

\subsection{The set-up}
The full vacuum state is the product of the vacuum state for each oscillator. Each oscillator is labeled by $p,\ell, m$. The Bunch-Davies vacuum is then defined as 
\begin{eqnarray}
|0\rangle_{\rm BD}=\prod_p |0_p\rangle_{\rm BD}\,,
\end{eqnarray}
and each mode is given by Eq.~(\ref{schmidt2}):
\begin{eqnarray}
|0_p\rangle_{\rm BD}&=&\sqrt{1-|\gamma_p|^2}\,\sum_{n=0}^\infty\gamma_p^n\,
|n_p\rangle_{R^\prime}|n_p\rangle_{L^\prime}\,,
\end{eqnarray}
where we omitted the indices $\ell, m$ of each oscillator for simplicity.

In the structure of the multiverse, there may be many causally disconnected de Sitter universes (de Sitter bubbles). Some of their quantum states may be far from the Bunch-Davies vacuum and be entangled with those of the other part of the multiverse as shown in~\cite{Maldacena:2012xp}. To model such a situation, we consider two modes, $p=k$ and $s$ of the free massive scalar field Eq.~(\ref{action}) in two de Sitter spaces in a maximally entangled pure state:
\begin{eqnarray}
|\psi\rangle=\frac{1}{\sqrt{2}}\Bigl(\,|0_s\rangle_{\rm BD1}|0_k\rangle_{\rm BD2}+|1_s\rangle_{\rm BD1}|1_k\rangle_{\rm BD2}\,\Bigr)\,,
\label{max}
\end{eqnarray}
where the states $|0_s\rangle_{\rm BD1}$ and $|1_s\rangle_{\rm BD1}$ are the vacuum and single particle excited states of the mode $s$ in a de Sitter space (BD1) and similarly for the other de Sitter space (BD2). We assume that the outside observer has a detector which only detects mode $s$ and the inside observer has a detector sensitive only to mode $k$.\footnote{The case of entanglement between an inertial and a noninertial frame for a free massless scalar field in Minkowski space is discussed in~ \cite{FuentesSchuller:2004xp}.}

Note that a quantum mechanical system here consist of subspaces BD1 and BD2.  The Hilbert space becomes a direct product ${\cal H}={\cal H}_{\rm BD1}\otimes{\cal H}_{\rm BD2}$.

\subsection{The single particle excitation state}

Let us introduce a $4\times4$ matrix form of Eq.~(\ref{bc}),
\begin{eqnarray}
c_J=b_I\,G^I{}_J\,,\qquad
G^I{}_J=\left(
\begin{array}{ll}
U^\sigma{}_{q} &~ V^{\sigma*}{}_{\!\!\!q} \\
V^\sigma{}_{q} &~ U^{\sigma*}{}_{\!\!\!q} \\
\end{array}\right)\,,\qquad
c_J=(c_{q}\,,c_{q}^\dagger)\,,
\label{bc1}
\end{eqnarray}
where $U^\sigma{}_{q}\equiv{\rm diag}(u,\bar{u})$, $V^\sigma{}_{q}\equiv{\rm diag}(v,\bar{v})$. Then from Eqs.~(\ref{ab0}), (\ref{ab}) and (\ref{bc1}), we find the relation between operators $a_q$ and $c_q$ is given by
\begin{eqnarray}
a_J=c_K\left(G^{-1}\right)^K{}_I\left(M^{-1}\right)^I{}_J\,,
\end{eqnarray}
where
\begin{eqnarray}
\left(G^{-1}\right)^K{}_I\left(M^{-1}\right)^I{}_J=\left(
\begin{array}{ll}
Q^\sigma{}_{q} &~ R^{\sigma*}{}_{\!\!\!q} \\
R^\sigma{}_{q} &~ Q^{\sigma*}{}_{\!\!\!q} \\
\end{array}\right)\,.
\end{eqnarray}
The components of the above matrix are given by
\begin{eqnarray}
Q^\sigma{}_{q}=\left(
\begin{array}{cc}
Au &~ -Bu+D^*v \\
-Bu+D^*v &~ Au \\
\end{array}\right)\,,\qquad
R^\sigma{}_{q}=\left(
\begin{array}{cc}
-Av &~ Bv-D^*u \\
Bv-D^*u &~ -Av \\
\end{array}\right)\,,
\end{eqnarray}
where we have used the relations
\begin{eqnarray}
A^*=A\,,\qquad B^*=-B\,,\qquad u^*=u=\bar{u}\,,\qquad v^*=v=\bar{v}\,.
\end{eqnarray}
Note that $B,\,v=0$ for the cases of conformal invariance ($\nu=1/2$) and masslessness ($\nu=3/2$).

The single particle excitation state of the inside observer is then calculated as
\begin{eqnarray}
|1_k\rangle_{\rm BD2}&=&a_L^\dagger\,|0_k\rangle_{\rm BD2}\\
&=&\left(Auc_L^\dagger-Avc_L+\left(Bu+Dv\right)c_R^\dagger-\left(Bv+Du\right)c_R\right)
|0_k\rangle_{\rm BD2}\nonumber\\
&=&f\sqrt{1-|\gamma_k|^2}\sum_{n=0}^\infty\gamma_k^n\sqrt{n+1}\,
|n_k\rangle_{R^\prime}|(n+1)_k\rangle_{L^\prime}\nonumber\\
&&\qquad+g\sqrt{1-|\gamma_k|^2}\sum_{n=0}^\infty\gamma_k^n\sqrt{n+1}\,
|(n+1)_k\rangle_{R^\prime}|n_k\rangle_{L^\prime}\,,\nonumber
\end{eqnarray}
where we used
\begin{eqnarray}
c^\dagger\,|n\rangle=\sqrt{n+1}\,|n+1\rangle\,,\quad
c\,|n\rangle=\sqrt{n}\,|n-1\rangle\,,
\end{eqnarray}
and defined 
\begin{eqnarray}
f=Au-\left(Bv+Du\right)\gamma_k\,,\quad g=-Av\,\gamma_k+Bu+Dv\,.
\end{eqnarray}
Since the inside observer remains constrained to the region $L^\prime$ in an de Sitter space, we shall compute the density matrix by tracing out the degree of freedom of region $R^\prime$ in the next subsection. This procedure corresponds to dividing the subspace BD2 into two more subspaces $R^\prime$ and $L^\prime$. The Hilbert space becomes a direct product ${\cal H}={\cal H}_{\rm BD1}\otimes{\cal H}_{R^\prime}\otimes{\cal H}_{L^\prime}$. 

\subsection{The density matrix of the inside observer}
The original maximally entangled state Eq.~(\ref{max}) is now expressed in terms of the Bunch-Davies modes for the outside observer and $R^\prime, L^\prime$ modes for the inside observer:
\begin{eqnarray}
|\psi\rangle&=&\frac{1}{\sqrt{2}}\,|0_s\rangle_{\rm BD1}\,\sqrt{1-|\gamma_k|^2}\,\sum_{n=0}^\infty\gamma_k^n\,|n_k\rangle_{R^\prime}|n_k\rangle_{L^\prime}
\nonumber\\
&&+\frac{1}{\sqrt{2}}|1_s\rangle_{\rm BD1}\left(
f\sqrt{1-|\gamma_k|^2}\,\sum_{n=0}^\infty\gamma_k^n\sqrt{n+1}\,
|n_k\rangle_{R^\prime}|(n+1)_k\rangle_{L^\prime}\right.\nonumber\\
&&\left.\hspace{2.5cm}+g\sqrt{1-|\gamma_k|^2}\,\sum_{n=0}^\infty\gamma_k^n \sqrt{n+1}\,
|(n+1)_k\rangle_{R^\prime}|n_k\rangle_{L^\prime}\right)\,.
\end{eqnarray}
Note that the scale dependence comes in the state $|\psi\rangle$ via $\gamma_k\,,f$ and $g$. This is because the Hilbert space of the inside observer was divided into two subspaces: ${\cal H}_{\rm BD2}={\cal H}_{R^\prime}\otimes{\cal H}_{L^\prime}$.

Since the inside observer is causally disconnected from region $R^\prime$, the observer must trace over the states in the region, which results in a mixed state
\begin{eqnarray}
\rho&=&{\rm Tr}_{R^\prime}|\psi\rangle\langle\psi|
=\sum_{m=0}^\infty{}_{R^\prime}\langle m|\psi\rangle\langle\psi|m\rangle_{R^\prime}
\nonumber\\
&=&\frac{1-|\gamma_k|^2}{2}\sum^\infty_{m=0}
|\gamma_k|^{2m}\,\rho_m\,,
\label{dm2}
\end{eqnarray}
where
\begin{eqnarray}
\rho_m&=&|0m\rangle\langle 0m|+g^*\gamma_k\sqrt{m+1}~|0m+1\rangle\langle 1m|
+g\gamma_k^*\sqrt{m+1}~|1m\rangle\langle 0m+1|
\nonumber\\
&&+|g|^2(m+1)~|1m\rangle\langle 1m|
%\nonumber\\
%&&
+f^*\sqrt{m+1}~|0m\rangle\langle 1\,m+1|+f\sqrt{m+1}~|1\,m+1\rangle\langle 0m|
\nonumber\\
&&+f^*g\gamma_k^*\sqrt{(m+1)(m+2)}~|1m\rangle\langle 1\,m+2|+fg^*\gamma_k\sqrt{(m+1)(m+2)}~|1\,m+2\rangle\langle 1m|
\nonumber\\
&&+|f|^2\,(m+1)~|1\,m+1\rangle\langle 1\,m+1|\,,
\end{eqnarray}
where $|nm\rangle=|n_s\rangle_{\rm BD1}|m_k\rangle_{L^\prime}$. Note that infinite degree of freedom labeled by $m$ comes in the state of the inside observer by confining to one of the regions of the open chart.

\subsection{The partial transpose and the negative eigenvalues}
\label{s4.4}

\begin{figure}[t]
\vspace{-1.5cm}
\includegraphics[height=10cm]{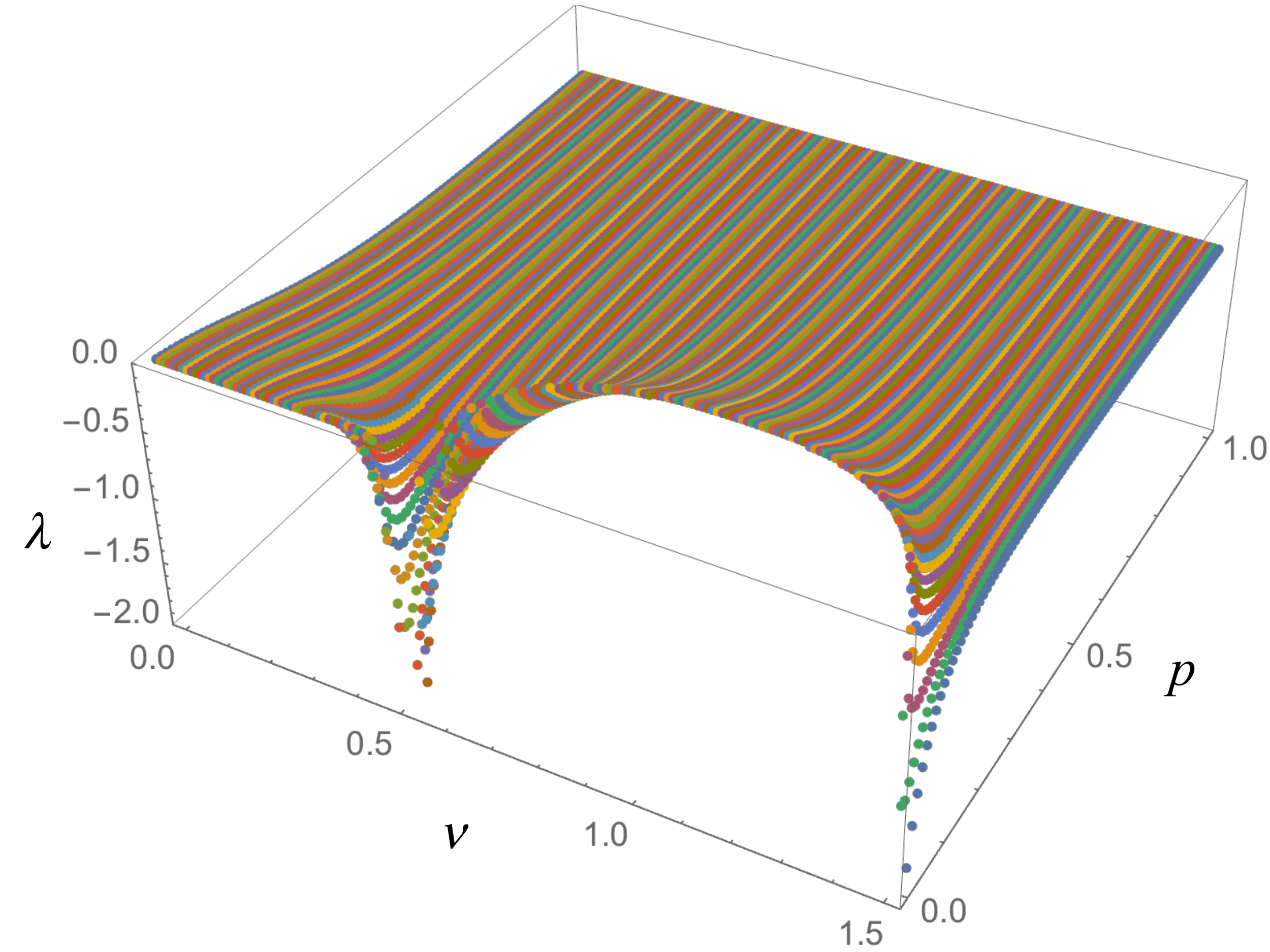}\centering
%\vspace{-0.5cm}
\caption{3D plot of the negative eigenvalues $\lambda$ as a function of $\nu$ and $p$.
We see that the negativity becomes large or small for small $p$ depending on the mass of the scalar field.}
\label{fig3}
\end{figure}

We obtain the partial transpose with respect to the subsystem ${\cal H}_{\rm BD1}$
\begin{eqnarray}
\rho_m^{T_{\rm BD1}} &=&|0m\rangle\langle 0m|+g^*\gamma_k\sqrt{m+1}~|1m+1\rangle\langle 0m|
+g\gamma_k^*\sqrt{m+1}~|0m\rangle\langle 1m+1|
\nonumber\\
&&
+|g|^2(m+1)~|1m\rangle\langle 1m|
%\nonumber\\
%&&
+f^*\sqrt{m+1}~|1m\rangle\langle 0\,m+1|+f\sqrt{m+1}~|0\,m+1\rangle\langle 1m|
\nonumber\\
&&+f^*g\gamma_k^*\sqrt{(m+1)(m+2)}~|1m\rangle\langle 1\,m+2|+fg^*\gamma_k\sqrt{(m+1)(m+2)}~|1\,m+2\rangle\langle 1m|
\nonumber\\
&&+|f|^2\,(m+1)~|1\,m+1\rangle\langle 1\,m+1|\,.
\label{pt1}
\end{eqnarray}
If at least one eigenvalue of the partial transpose is negative, then the density matrix is entangled and the state between inside and outside observers is entangled. We compute the eigenvalues $\lambda$ numerically and the resultant negative eigenvalues are plotted in Figure~\ref{fig3} for $0<\nu<1.5$ and $0<p<1$.  Because larger negative eigenvalues means stronger entanglement, we can read off that the entanglement gets stronger as we go to the large scale ($p\rightarrow 0$) when the mass of scalar field is around massless ($\nu=3/2$) and conformally invariant ($\nu=1/2$). We also see that the entanglement vanishes in a region centered at $\nu=1$ on large scales, but other than that, the state remains correlated. In Figure~\ref{fig4}, we plot the slices of $p, \nu={\rm constant}$ of the negative eigenvalues separately. From the right panel, we find that the negative eigenvalues for $\nu\sim1/2,~3/2$ start to increase around $3$ times the curvature scale of the open universe ($p\sim 0.3$). This is consistent with the result in \cite{Kanno:2014ifa} where the entanglement affects the shape of the spectrum on large scales comparable to or greater than the curvature radius when the mass of the scalar field is $m^2=H^2/10~(\nu\sim1.47)$. On the other hand, the entanglement vanishes for a finite $p$ when $\nu$ is fixed in the interval $1/2<\nu<3/2$. In the case of entanglement between an inertial and a noninertial frame in Minkowski space discussed in~\cite{FuentesSchuller:2004xp}, the entanglement vanishes only in the high acceleration limit. Thus, this result should be the specific to de Sitter space. The vanishing negativity in the $(p, \nu)$ plane is found in Figure~\ref{fig4-2}.

\subsection{Negativity in the small scale limit}

\begin{figure}[t]
\begin{center}
\vspace{-1cm}
\begin{minipage}{8.1cm}
\includegraphics[height=6cm]{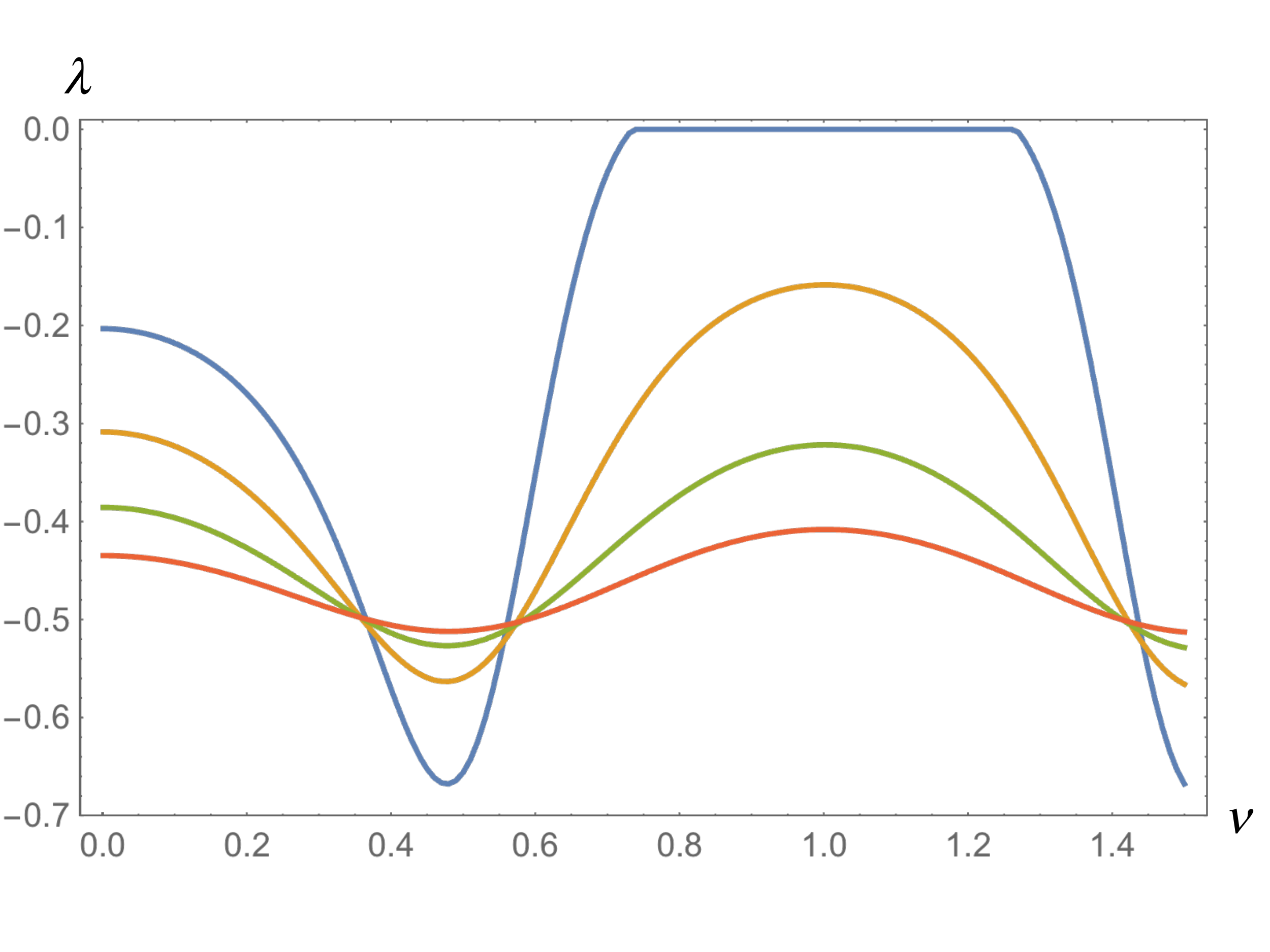}\centering
\end{minipage}
\begin{minipage}{8.1cm}
\hspace{0.5cm}
\includegraphics[height=6cm]{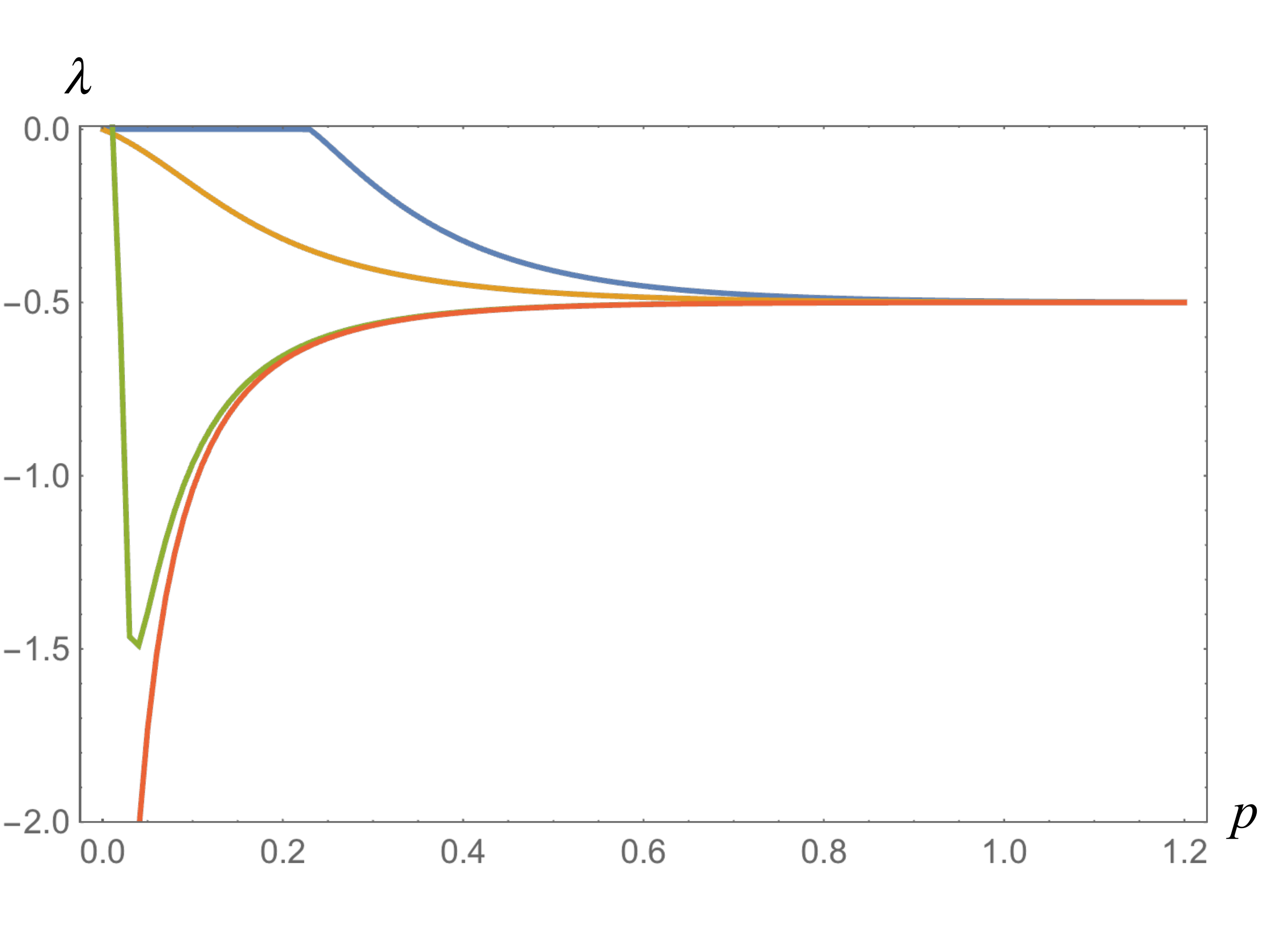}
\end{minipage}
\caption{The left panel shows plots of the sum of negative eigenvalues $\lambda$ versus $\nu$. The red line is for $p=0.5$, the green is for 0.4, the yellow is for 0.3 and the blue is for 0.2. The right panel shows plots of the most negative eigenvalues $\lambda$ versus $p$. The red line is for $\nu=0.5,~1.5$, the green is for $0.49$, the yellow is for $0.25$ and the blue is for $1.0$.}
\label{fig4}
\end{center}
\end{figure}

Let us examine the negativity in the small scale limit $p\rightarrow\infty$. In this limit, we find $f\rightarrow 1$ and $g\rightarrow 0$, so the partial transpose Eq.~(\ref{pt1}) becomes 
\begin{eqnarray}
\rho_m^{T_{\rm BD1}} &=&|0m\rangle\langle 0m|+\sqrt{m+1}~|1m\rangle\langle 0\,m+1|+\sqrt{m+1}~|0\,m+1\rangle\langle 1m|
\nonumber\\
&&+(m+1)~|1\,m+1\rangle\langle 1\,m+1|\,.
\end{eqnarray}
By plugging this back in Eq.~(\ref{dm2}), we see that only the $m=0$ term remains in the small scale limit due to the fact that $|\gamma_k|\rightarrow 0$ and the negative eigenvalue is found to be $-1/2$. Because the initially maximally entangled state in Eq.~(\ref{max}) also gives the same negative eigenvalue $-1/2$, this means that the initial state of entanglement is recovered in the small scale limit. We can see this in the right panel of Figure~\ref{fig4}.

In subsection \ref{s4.4}, we found that the entanglement on large scales became stronger or weaker than that on small scales depending on the mass of the scalar field. So, this means that the entanglement gets stronger or weaker on large scales than that of initially maximally entangled state. Mathematically, the reason for more entanglement is that infinite degree of freedom of the state comes in the inside observer's state by confining to one of the regions of the open chart as in Eq.~(\ref{dm2}). Thus, we could say that the increase of entanglement is due to the particular point of view of the observer and that the quantum entanglement is thus observer dependent. The reason for getting less entanglement and eventually no entanglement  would be similar to the case of an accelerated observer in Minkowski space discussed in~\cite{FuentesSchuller:2004xp}, which shows that the entanglement is an observer dependent quantity.

\begin{figure}[t]
\vspace{-1.5cm}
\includegraphics[height=8cm]{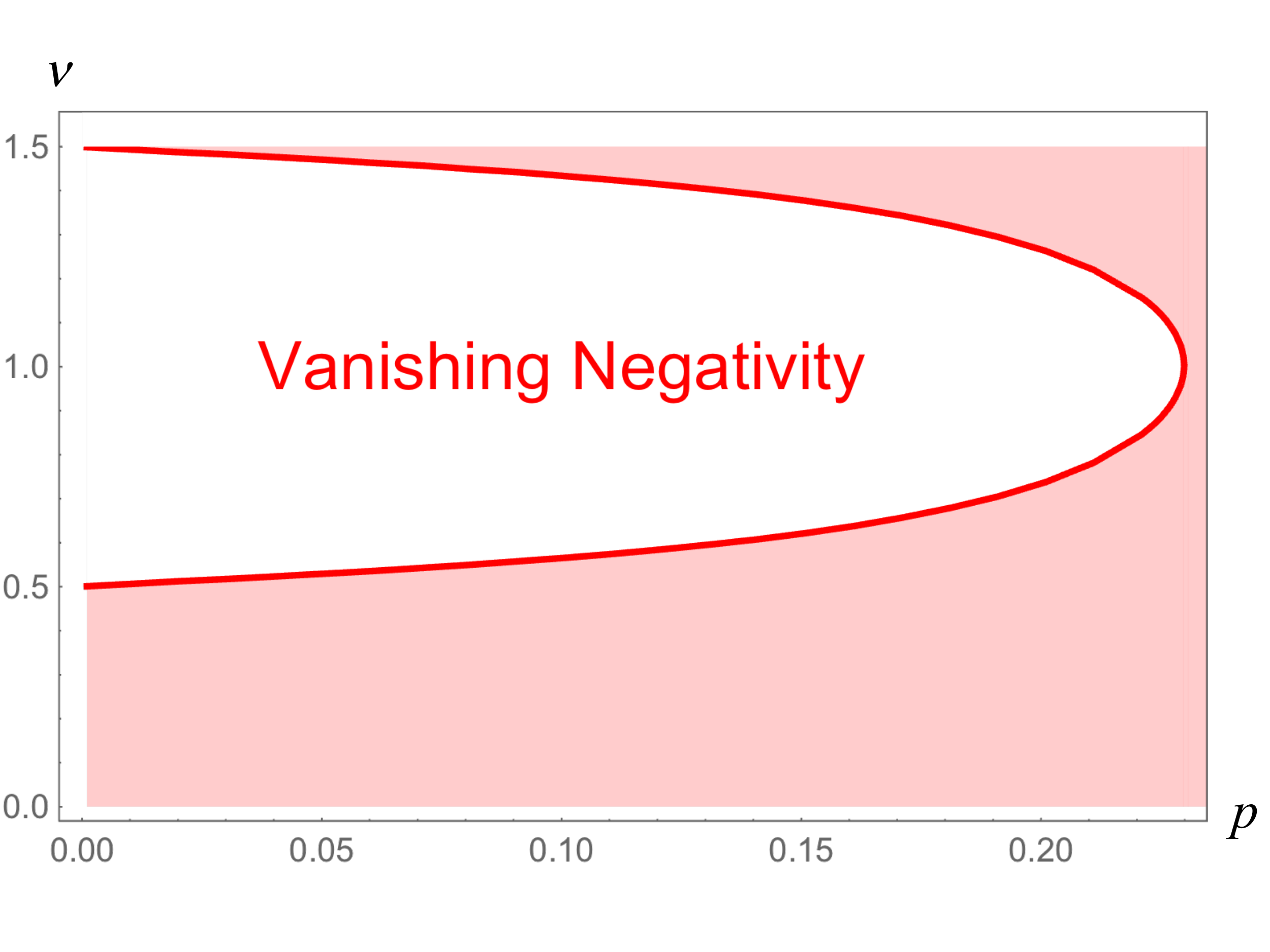}\centering
\vspace{-1cm}
\caption{The region of vanishing negativity.}
\label{fig4-2}
\end{figure}

\subsection{Negativity in the massless limit}

In the cases of conformal invariance ($\nu=1/2$) and masslessness ($\nu=3/2$), we find $f\rightarrow\left(A-D\gamma_k\right)u$ and $g\rightarrow 0$. Then the partial transpose Eq.~(\ref{pt1}) becomes 
\begin{eqnarray}
\rho_m^{T_{\rm BD1}} &=&|0m\rangle\langle 0m|+f^*\sqrt{m+1}~|1m\rangle\langle 0\,m+1|+f\sqrt{m+1}~|0\,m+1\rangle\langle 1m|
\nonumber\\
&&+|f|^2\,(m+1)~|1\,m+1\rangle\langle 1\,m+1|\,,
\end{eqnarray}
where
\begin{eqnarray}
f=\left(A-D\gamma_k\right)u=\frac{1}{2\sinh\pi p}\left(e^{\pi p}-i\,e^{-\pi p}\,\frac{1+ip}{1-ip}\frac{\Gamma\left(ip\right)}{\Gamma\left(-ip\right)}\right)\,.
\end{eqnarray}
As shown in Eq.~(\ref{thermal}), this should reflect in the thermal nature of the state.

In this simple case, we find the negative eigenvalues in the $(m, m+1)$ block to be
\begin{eqnarray}
\lambda_m=\frac{|\gamma_k|^{2m}\left(1-|\gamma_k|^2\right)}{4} \left(\,\frac{m|f|^2}{|\gamma_k|^2}+|\gamma_k|^2 
 - \sqrt{\left(\frac{m|f|^2}{|\gamma_k|^2}+|\gamma_k|^2 \right)^2+4|f|^2}\,\right)\,.
\end{eqnarray}
Note that the eigenvalues in this case are very similar to those in the case of entanglement between an inertial and a noninertial frame for a free massless scalar field in Minkowski space discussed in~ \cite{FuentesSchuller:2004xp}, but are never identical. Since the result of~ \cite{FuentesSchuller:2004xp} accounts for the thermal property of Minkowski space, the difference from it here should be due to the property of de Sitter expansion.

\begin{figure}[t]
\vspace{-2.5cm}
\includegraphics[height=10cm]{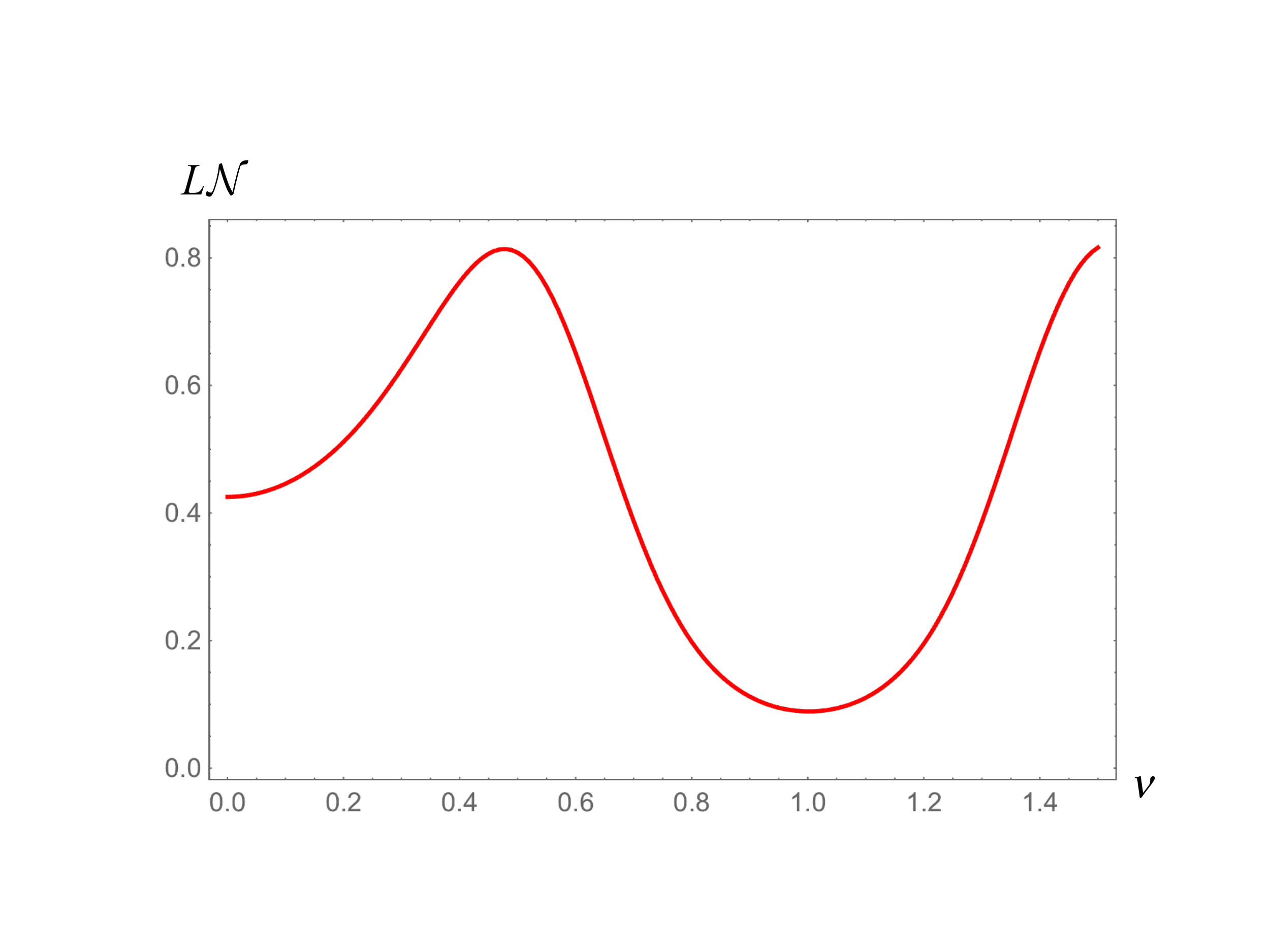}\centering
\vspace{-1cm}
\caption{Plot of the logarithmic negativity as a function of $\nu$ for $p=0.25$.}
\label{fig5}
\end{figure}

\subsection{The logarithmic negativity}

Now, we sum over all the negative eigenvalues and calculate the logarithmic negativity defined in Eq.~(\ref{lnvsn}). Since we focus on an observer's detector for modes of momentum $p$, we don't integrate either over $p$ nor a volume integral over the hyperboloid. The result is found in  Figure~\ref{fig5} where we take $p=0.25$. The qualitative feature of it is similar to Figure~\ref{fig2}. In both cases of entanglement between two causally disconnected open charts in a de Sitter space and two causally separated de Sitter spaces, two peaks appear, one for the conformally invariant case ($\nu=1/2$) and one for the massless case ($\nu=3/2$). These had in fact appeared in the negative eigenvalues at small $p$ in the left panel of Figure~\ref{fig4}. Thus, if we take $p\sim 0.1$, the logarithmic negativity becomes similar to the blue line in the left panel of Figure~\ref{fig4} and the state is no longer entangled in a region centered at $\nu=1.0$ as seen in Figure~\ref{fig4-2}. This would reflect the property of de Sitter expansion analogous to the case of the high acceleration limit of an accelerated observer in Minkowski space~\cite{FuentesSchuller:2004xp}.

\section{Summary and discussion}
\label{s5}

In this work we have studied the entanglement negativity of a free massive scalar field between two causally disconnected regions of the multiverse.
Firstly, we calculated the negativity between two causally disconnected open charts in de Sitter space. Since the inside of a nucleated bubble is known to look like an open universe,
this setup corresponds to studying the entanglement between inside and outside of a simplified bubble without a bubble wall. We found that the qualitative feature of it agrees with the one calculated by entanglement entropy in~\cite{Maldacena:2012xp}, that is, the entanglement as a function of mass parameter squared has two peaks when the scalar field is conformally invariant and massless, and for large mass the entanglement decays exponentially. The oscillating behavior for small mass should be related to the balance between the mass of the scalar field and de Sitter expansion, but this remains an open question.

We then introduced two observers who determine the entanglement between two causally separated de Sitter spaces, supposing that they are initially in a maximally entangled pure state in the structure of the multiverse. We set one of the observers to remain inside of a simplified bubble, that is, remaining constrained to the region $L$ of the open chart in a de Sitter space and set the other observer to be the other de Sitter space. We computed the negativity in this setup and found that the scale dependence enters into the entanglement when the Hilbert space of the inside observer was divided into two subspaces of $R$ and $L$ (${\cal H}_{\rm BD2}={\cal H}_{R}\otimes{\cal H}_{L}$). We showed that the initially maximally entangled state becomes more or less entangled on large scales depending on the mass of the scalar field and recovers the initially entangled state in the small scale limit. Mathematically, the increase of entanglement is because the infinite degrees of freedom of the state come in the inside the observer's state by confining to one of the regions of the open chart. So we found that the increase of entanglement is due to the observer's point of view and that the quantum entanglement is observer dependent. The reason for the decrease in entanglement would be similar to the case of an accelerated observer in Minkowski space as discussed in~\cite{FuentesSchuller:2004xp}. This reflects that the entanglement is an observer dependent quantity as well. We also showed that the entanglement remains on large scales when the scalar field is close to the cases of masslessness or conformal invariance. This result is consistent with the one obtained in~\cite{Kanno:2014ifa} where the entanglement affects the shape of the spectrum on large scales comparable to or greater than the curvature scale of the open universe when the scalar mass is $m^2=H^2/10~(\nu\sim1.47)$. We then calculated the logarithmic negativity and showed that the result, as a function of mass parameter for a fixed mode around $p\sim 0.3$, is similar to the entanglement between two causally separated open charts in de Sitter space. On larger scales $p\sim 0.1$, the entanglement vanishes in a region centered at $\nu=1$. We speculated that this reflects the property of the de Sitter expansion analogous to the case of the high acceleration limit of an observer in Minkowski space, but this remains another open question. We note that the negativity does not vanish even in the small scale limit ($p\rightarrow\infty$).
It would be interesting  to see if  the effect of entanglement on small scales can be observed. Indeed, it might appear in the initial state as a non-Bunch Davies vacuum~\cite{Kanno:2014lma, Albrecht:2014aga}.

We also discussed the cases of the conformally invariant and massless scalar, which turned out to have a thermal property. This is because the density matrix becomes a thermal state with temperature $H/(2\pi)$. This reminds us of the phenomenon of a consequence of the Unruh effect by an accelerated observer, and confirms that the entanglement is an observer dependent quantity in de Sitter space as well. In the study of~\cite{Kanno:2014ifa}, the difference in the spectra of vacuum fluctuations between the originally entangled state and the mixed state after tracing out the inaccessible region (the difference between blue and red lines) disappeared in these cases. So these cases might relate to some peculiar property of de Sitter space. It would be interesting to examine those cases in more details.

It would also be interesting to make use of the entanglement of primordial gravitational waves to find observational signatures in the multiverse, as we found that massless scalar field tends to increase the entanglement on large scales. Also gravitational waves are less interactive and the entanglement of them could carry the information of the multiverse without possible contaminations.

From the above arguments, we expect that quantum entanglement would be able to provide some evidence for the existence of the multiverse.

\section*{Acknowledgments}
We would like to thank Yasusada Nambu for fruitful discussions.
This work was supported by IKERBASQUE, the Basque Foundation for
Science, Grants-in-Aid for Scientific Research (C) No.25400251 and Grants-in-Aid for Scientific Research on Innovative Areas No.26104708. 

%
%\appendix

\end{document}